\documentclass[12pt]{spieman}  
\usepackage{amsmath,amsfonts,amssymb}
\usepackage{graphicx}
\usepackage{setspace}
\usepackage{tocloft}

\usepackage{times}
\usepackage{soul}
\usepackage{graphicx}
\usepackage{amsmath}
\usepackage{booktabs}
\usepackage{amssymb}
\usepackage{latexsym}
\usepackage{tikz}
\usepackage{url}
\usepackage{xcolor}
\usepackage{array}
\usepackage{multirow}
\usepackage{algorithm}
\usepackage{algorithmic}
\usepackage{enumitem}
\usepackage{mathtools}
\usepackage{hyperref}
\usepackage{changepage}
\usepackage{amssymb}
\usepackage{graphicx}
\usepackage{textcomp}
\usepackage{booktabs}
\usepackage{color}
\usepackage{cite}
\usepackage{setspace}

\title{Constrained Deep One-Class Feature Learning For Classifying Imbalanced Medical Images}

\author[a,b]{Long Gao}
\author[c]{Chang~Liu}
\author[a]{Dooman~Arefan}
\author[d]{Ashok Panigrahy}
\author[a,c,e,f*]{Shandong~Wu}
\affil[a]{Department of Radiology, School of Medicine, University of Pittsburgh, 4200 Fifth Ave, Pittsburgh PA, USA, 15260}
\affil[b]{College of Computer, National University of Defense Technology, Changsha, 410073}
\affil[c]{Department of Bioengineering, Swanson School of Engineering, University of Pittsburgh, 4200 Fifth Ave, Pittsburgh PA, USA, 15260}
\affil[d]{Department of Radiology,  University of Pittsburgh Medical Center Children's Hospital of Pittsburgh, 4401 Penn Ave, Pittsburgh, PA, USA, 15224}
\affil[e]{Department of Biomedical Informatics, University of Pittsburgh, 4200 Fifth Ave, Pittsburgh, USA, 15260}
\affil[f]{Intelligent Systems Program, University of Pittsburgh, 4200 Fifth Ave, Pittsburgh PA, USA, 15260}

\cftpagenumbersoff{figure}
\cftpagenumbersoff{table} 
\begin{document} 
\maketitle

{\noindent \footnotesize\textbf{*Corresponding author:} Shandong Wu,  \linkable{wus3@upmc.edu}}

\begin{abstract}

\begin{spacing}{2}
Medical image data are usually imbalanced across different classes. One-class classification has attracted increasing attention to address the data imbalance problem by distinguishing the samples of the minority class from the majority class. Previous methods generally aim to either learn a new feature space to map training samples together or to fit training samples by autoencoder-like models. These methods mainly focus on capturing either compact or descriptive features, where the information of the samples of a given one class is not sufficiently utilized.
In this paper, we propose a novel deep learning-based method to learn compact features by adding constraints on the bottleneck features, and to preserve descriptive features by training an autoencoder at the same time. Through jointly optimizing the constraining loss and the autoencoder’s reconstruction loss, our method can learn more relevant features associated with the given class, making the majority and minority samples more distinguishable. 
Experimental results on three clinical datasets (including the MRI breast images, FFDM breast images and chest X-ray images) obtain state-of-art performance compared to previous methods.
\end{spacing}


\end{abstract}

\keywords{Data imbalance, medical imaging, deep learning, one-class classification}

\begin{spacing}{2}   

\section{Introduction}

In clinical and biomedical applications, the collected datasets are usually imbalanced across different classes. Models trained on imbalanced data may lead to poor or biased performance. To address this issue, One-Class Classification (OCC) $^{[}$\cite{khan2014one}$^{]}$ has been proposed to learn single-class-relevant features from samples belonging to the "majority" class. This method has also been used for Anomaly Detection (AD) $^{[}$\cite{chalapathy2019deep,ieracitano2019convolutional}$^{]}$, which focuses on distinguishing the "minority" class samples that have different distribution from the "majority" class.

The critical problem for OCC is how to learn a high-quality feature space. A good feature space should be both compact and descriptive $^{[}$\cite{2018Learning,Tax2004svdd}$^{]}$. The compactness means the model can learn a similar representation for different images belonging to the same class. In this way, samples from the same class should have lower intra-class distance. The descriptiveness means the model can learn different representations for images belonging to different classes. In this way, samples from different classes should have higher inter-class distance. 
To optimize the compactness, the model should map images from the same class to similar representations. On the other hand, to optimize the descriptiveness, the model should capture features that belong exclusively to each class. Considering that abnormal samples are unavailable during the training phase in OCC, we focus on learning more features of the given class to optimize the descriptiveness. Based on the feature that the model focuses on, previous methods can be generally split into two categories: learning compact features or learning descriptive features. 

The first scheme is to map features by kernel functions. These methods usually focus on learning a compact representation by minimizing the intra-class distance.
A representative method is OCSVM $^{[}$\cite{scholkopf2000support}$^{]}$, which focuses on mapping given class samples from the original space to a new feature space and maximizing the distance between the original point and the hyperplane. SVDD $^{[}$\cite{Tax2004svdd}$^{]}$ attempts to learn a kernel function, where the given class samples can be mapped into a hypersphere, and samples belonging to the other classes will be mapped out of it. Some work also applies these methods to clinical imaging analysis $^{[}$\cite{chen2018diverse,dreiseitl2010outlier,guo2011tumor}$^{]}$. These methods obtain remarkable performance for small-scale datasets because the kernel function can map samples into a separable higher dimensional feature space effectively, but they usually suffer from the "curse of dimensionality" $^{[}$\cite{ruff18deepsvdd}$^{]}$.  To address this issue, D-SVDD $^{[}$\cite{ruff18deepsvdd}$^{]}$ is proposed to learn a hypersphere that encloses the network representations of the data by an encoder network. These methods can learn a compact representation for the given class samples, but they neglect the descriptive features of the images. Thus, these models may learn a trivial solution such as mapping all samples into zeros.

Another scheme is to fit the distribution of the given class samples by deep learning models. These methods usually focus on learning descriptive features by reconstructing images. Deep neural networks have been widely employed in this scheme because of their powerful feature learning abilities $^{[}$\cite{yosinski2014transferable,zong2018deep}$^{]}$. Example methods include convolutional autoencoders (CAE) $^{[}$\cite{deecke2018anomaly,zhou2017anomaly}$^{]}$ and Generative Adversarial Network (GAN) based models $^{[}$\cite{Goodfellow2014Generative,tang2019chest,schlegl2017unsupervised}$^{]}$. For example, Deep Autoencoding Gaussian Mixture Model (DAGMM) $^{[}$\cite{fan2020robust,zhou2017anomaly}$^{]}$ is proposed to add a Gaussian Mixture Model on the output layer to learn more local features. AnoGAN $^{[}$\cite{deecke2018anomaly}$^{]}$ attempts to generate fake samples with GAN and takes the discriminator as the classifier to classify samples. Entropy-based loss is added in the Deep Structured Energy-Based Model (DSEBM) $^{[}$\cite{zhou2017anomaly}$^{]}$ to make samples belonging to different classes more distinguishable. The study in $^{[}$\cite{wei2018anomaly}$^{]}$ employs an inpainting method to detect tumors by calculating the reconstruction loss of each patch. These methods usually focus on reconstructing input features and can learn a descriptive model. However, they usually neglect the compactness of the feature space. Thus, the performance of these algorithms is limited, especially on clinical image datasets with large variations.

In this paper, a novel method is proposed to learn a compact and descriptive hypersphere using the autoencoder backbone. The intuition is that ideally, for a compact representation, all training samples should be mapped into a single point because they belong to the same class. However, models (such as Deep SVDD) that only use the encoder part of a network are easy to collapse to a trivial solution $^{[}$\cite{ruff18deepsvdd}$^{]}$ (e.g., map to all zero vectors). With that in mind, we present a new framework, namely $ConOCC$, to learn a more robust feature representation by combining a constraining layer and a decoder. The proposed method not only learns a compact hypersphere by using the constraining layer to pull all training samples together, but also preserves the descriptive information to avoid the trivial solution by reconstructing samples with a decoder. By jointly optimizing the constraining loss and reconstruction loss, the proposed method can obtain a balance between the compactness and descriptiveness of features to achieve better performance.

Our contributions can be summarized as follows: 
\begin{itemize}
    \item  A novel framework is proposed to learn a representative hypersphere for the one-class classification task by adding constraints on the features of the samples from the majority class.
    
    \item  The proposed method can learn a more compact and descriptive feature space by jointly optimizing two loss objectives.

    \item Extensive experiments are performed on three clinical datasets to demonstrate that the proposed method outperforms the state-of-the-art algorithms.
\end{itemize}

\section{Materials and Methods}

$\textbf{Definition}$ OCC is to learn a model $M$ from a set of given samples $X$ belonging to a single class $c$. A score $s_i$ will be obtained for each input sample $x_i$ by the model, where a lower score means $x_i$ is more likely to belong to the given class.

$\textbf{Motivation}$ Different from the classification of natural images, the classification of clinical images is usually more depend on the contrast and brightness features, making the classification of medical images more sensitive to the reconstruction loss of autoencoder-like models. 

Autoencoder has shown powerful ability in learning deep features from medical images $^{[}$\cite{tschannen2018recent}$^{]}$, and it has been widely applied in AD tasks $^{[}$\cite{chalapathy2019deep,zong2018deep}$^{]}$. To learn a more distinguishable hypersphere and to improve the performance, we add constraints on the bottleneck features. Jointly minimizing the hypersphere and reconstruction loss will force the encoder to extract the most representative features of a given class, because the network needs to closely map normal samples to the center of the hypersphere while preserving the information of normal samples.   The rationale of ConOCC is that, samples from the same class should ideally be mapped into the same point because they belong to the same class. In this way, we design ConOCC to pull all training samples to the center of the hypersphere in the feature space, as described in the following.

\subsection{Pipeline}  

\begin{figure*}[htbp]
\centering
\includegraphics[trim=100 360 40 80, scale=0.8]{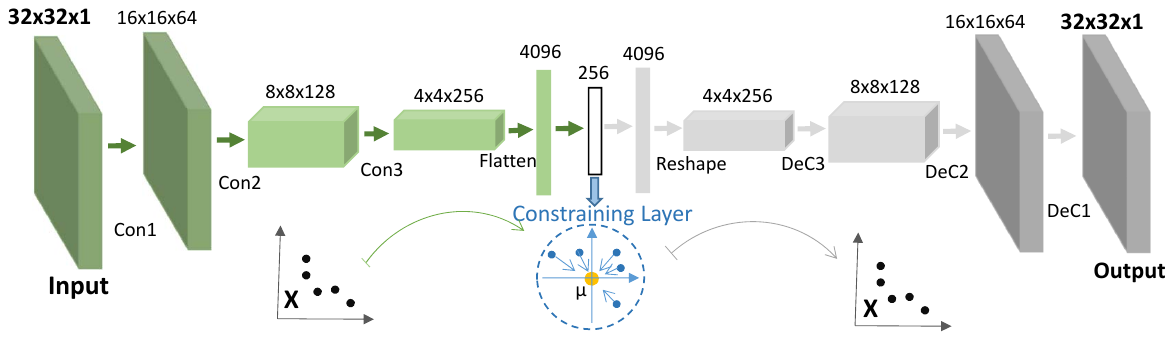}
\caption{\label{fig:structure} The framework of the proposed method. A constraining layer is added as a branch for the bottleneck feature to pull all features towards the center point $\mu$, as shown in the dotted circle. By jointly training the autoencoder and the constraining layer, the model can learn a compact while descriptive hypersphere. Samples from the majority class will be mapped into this hypersphere, while samples from the minority class will be mapped out of it.}
\end{figure*}

The framework of the proposed method is illustrated in Figure~\ref{fig:structure}. It includes an encoder $E$ (green), a decoder $D$ (grey), and a $Constraining$ $Layer$ (black). 

The function of each component in ConOCC is as follows: The $encoder$ can reduce the dimension and learn a basic hypersphere from the majority class samples. The $decoder$ can amplify the difference of features and prevent the model from mapping all samples into a trivial solution hypersphere (e.g., all zeros). The $Constraining\ Layer$ can minimize the distance between features and therefore compress the hypersphere. By selecting appropriate parameters, the model can obtain a balance between the constraining loss and the reconstruction loss, thus achieving a better performance. 

Suppose $\mu$ is the center point of the hypersphere, $z_i$ is the bottleneck feature generated by an encoder $E$ from the input sample $x_i$ of size $m\times m$. The method focuses on minimizing the distance between $z_i$ and $\mu$. The loss objective of the constraining layer can be defined as:

\begin{equation}
\label{equ:calc_q}
L_{con} = ||z_i-\mu||^2=||E(x_i)-\mu||^2
\end{equation}

Here Mean Square Error (MSE) is employed for the reconstruction loss as defined in \ref{mseloss}:

\begin{equation}
\label{mseloss}
L_{AE}=\sum_{j=1}^{j=m\times m}||D(E(x_i)) - x_i||_2^2
\end{equation}

To combine the constraining loss and the reconstruction loss together, the loss function of the whole network can be formulated as:

\begin{equation}
L = L_{AE} + \gamma L_{con}
\end{equation}

where $\gamma$ is the coefficient to balance the influence of $L_{AE}$ and $L_{con}$. $L_{AE}$ is the reconstruction loss of the autoencoder.

Compared to D-SVDD which penalizes the trivial solution $^{[}$\cite{ruff18deepsvdd}$^{]}$, the decoder employed in ConOCC not only avoids the trivial solution by reconstructing inputs, but also promotes the encoder to learn descriptive features of the given class.  

Compared to Sparse Autoencoder $^{[}$\cite{glorot2011deep}$^{]}$ which also focuses on adding constraints on bottleneck features, ConOCC aims to minimize the distance of each feature node, instead of learning sparse representations.

\subsection{Optimization}

The weights of the encoder and the decoder are jointly updated with the constraining loss and the reconstruction loss. Specifically, the weights of the decoder $W_d$ are updated by:

\begin{equation}
W_d = W_d-\frac{\lambda}{b}\sum_{i=1}^{b}\frac{\partial L_{AE}}{\partial W_d} 
\end{equation}

where $\lambda$ and $b$ are the learning rate and mini-batch size, respectively. 
The weights of the encoder are updated by:

\begin{equation}
W_e = W_e-\frac{\lambda}{b}\sum_{i=1}^{b}(\frac{\partial L_{AE}}{\partial W_d} +\gamma \frac{\partial L_{con}}{\partial z_i})
\end{equation}

The center of the hypersphere $\mu$ is set to the mean value of the bottleneck feature. To stabilize the training procedure, we update $\mu$ after each $T$ epoch of training the whole model.

\subsection{Score calculation}

We use the score of each sample to evaluate how likely each of them belongs to the given class. The lower the score, the more likely sample $x_i$ is to belong to the given majority class. In this study, the reconstruction loss is employed as the score. The features extracted from samples of the majority class will result in a low reconstruction loss, while that of the minority class will result in a high reconstruction loss. In this way, the model can distinguish samples from different classes. Specifically, the score of sample $x_i$ can be calculated by the formula:

\begin{equation}
\label{equ:score1}
s_i= \sum_{j=1}^{j=m\times m}||D(E(x_i))-x_i||_2^2
\end{equation}

\subsection{Datasets}
\label{sec:dataset}

\begin{table*}[htbp]
\centering
\caption{Number of samples of each dataset. The "majority" class samples are used for training. We use 2-,5- and 4-fold cross-validation for the MRI, FFDM and ChestX datasets, respectively.}
\label{tab:data}
\begin{tabular}{p{1.5cm}<{\centering}|p{1.5cm}<{\centering}  p{2.2cm}<{\centering} p{1.5cm}<{\centering} p{2cm}<{\centering} p{1.2cm}<{\centering} p{1.2cm}<{\centering}}

\toprule[0.6pt]
\multirow{2}{*}{{Dataset}} &\multirow{2}{*}{{Dimension}} &\multicolumn{1}{c}{{Training}} &\multicolumn{1}{c}{{Majority}} &\multicolumn{1}{c}{{Minority}} & \multicolumn{1}{c}{{Testing}} &\multicolumn{1}{c}{{Cross}} \\
& &(Majority)  & in testing & in testing& in total & validation\\

\hline
MRI&32$\times$32$\times$1&1000& 946&926&1872&2\\
\hline
FFDM&64$\times$64$\times$1&200 &52&52&104&5\\
\hline
ChestX&100$\times$100$\times$1&1341 &234&390&624&4\\
\bottomrule[0.6pt]
\end{tabular}
\end{table*}

\begin{figure}[ht!]
\centering
\includegraphics[scale=0.4]{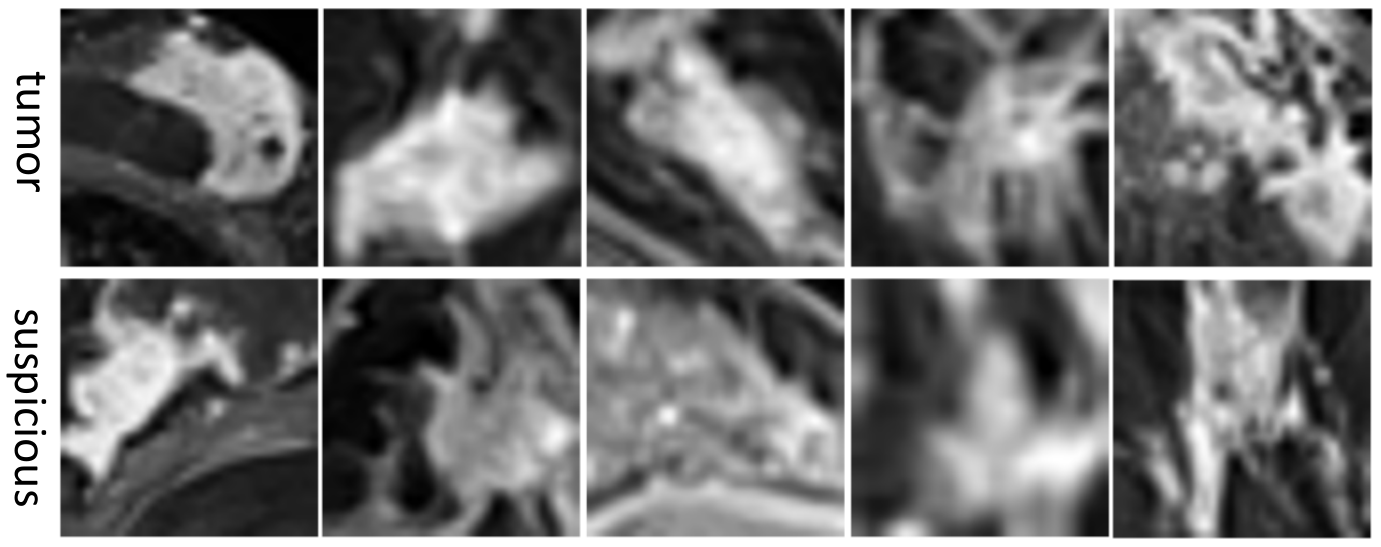}
\caption{\label{fig:MRI} Tumor (majority class) and suspicious (minority class) samples on the breast tumor MRI dataset.}
\end{figure}

\begin{figure}[ht!]
\centering
\includegraphics[scale=0.4]{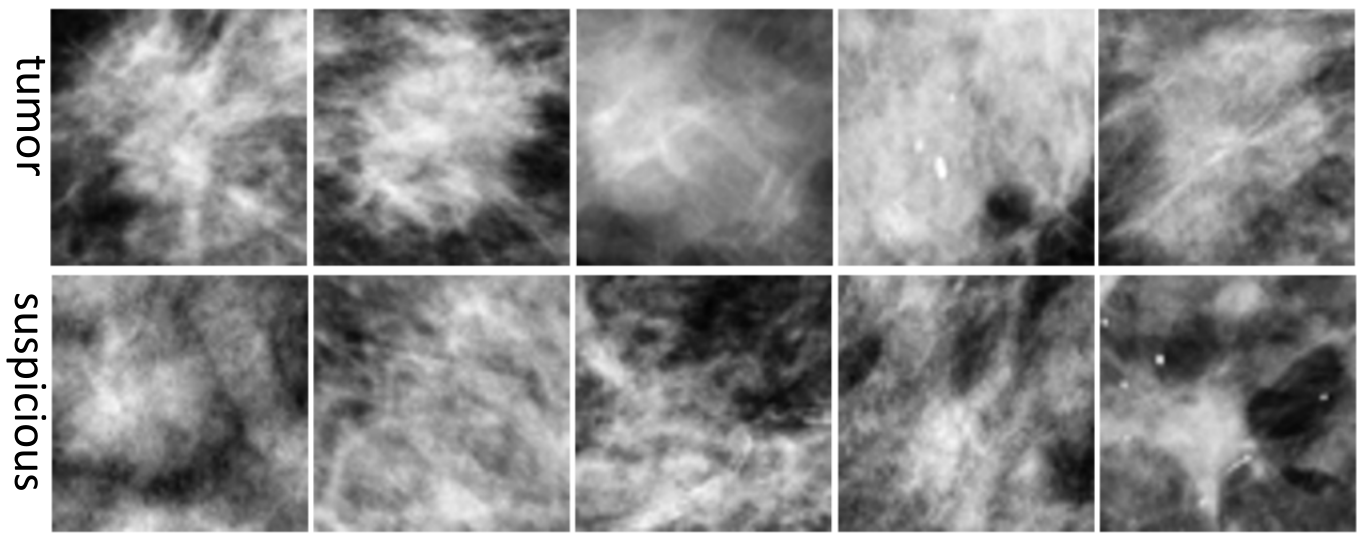}
\caption{\label{fig:ffdm} Tumor (majority class) and suspicious (minority class) samples on the breast tumor FFDM dataset.}
\end{figure}

\begin{figure}[ht!]
\centering
\includegraphics[scale=0.42]{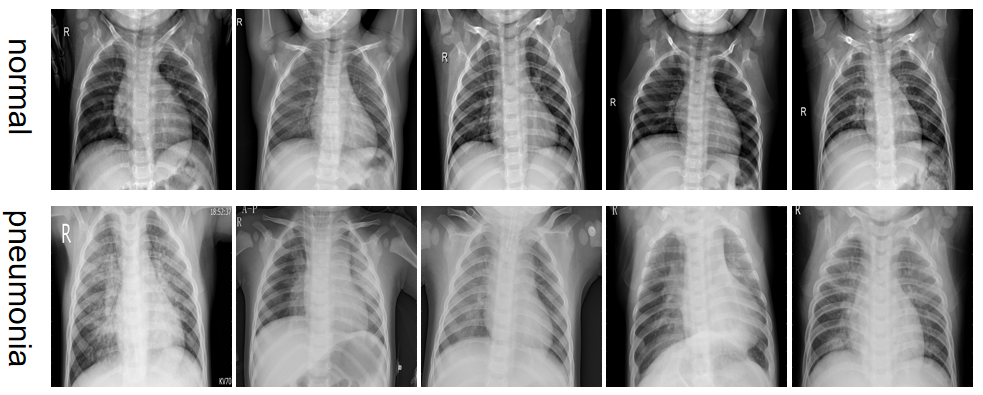}
\caption{\label{fig:ct} Normal (majority class) and pneumonia (minority class) samples on the ChestX dataset.}
\end{figure}

Three datasets from three modalities are employed for two clinical tasks: breast tumor detection and pneumonia diagnosis. The sample number of each dataset is shown in Table~\ref{tab:data}, some examples from the majority and minority classes are shown in Figures~\ref{fig:MRI},~\ref{fig:ffdm} and ~\ref{fig:ct}, and details are as follows.

$\textbf{Breast Magnetic Resonance Imaging (MRI) dataset}$: This is an MRI dataset collected at our institution from women diagnosed with breast cancer. 1000 tumor images (majority class) are used for training, and 946 tumor images and 926 normal (suspicious) tissue images are used for testing. As shown in Figure~\ref{fig:MRI}, the suspicious patches where tissue biopsy is needed are manually segmented from original MRI images. For each patient, one tumor/suspicious patch is used. The mean size of the tumor is about 31$\times$31 pixels. Here all the tumor patches are resized to 32$\times$32 pixels for all experiments.

$\textbf{Breast Full-Field Digital Mammography (FFDM)}$ $\textbf{dataset}$: This is an in-house X-ray breast imaging dataset for women diagnosed with breast cancer. 200 tumor (majority class) images are used for training, 52 tumor images and 52 normal (suspicious) tissue images are used for testing. The patch size ranges from 100 to 500 pixels, all the patches are resized to 64$\times$64 pixels by sub-sampling (see Figure~\ref{fig:ffdm}).

$\textbf{ChestX}$ $\textbf{dataset}$: This is a public chest X-rays dataset from Kaggle$\footnote{ https://www.kaggle.com/paultimothymooney/chest-xray-pneumonia
}$ for pneumonia diagnosis. It contains two classes: pneumonia patient samples and normal samples. 1341 normal CT images are used for training. 234 normal and 390 pneumonia images are used for testing. The original image size ranges from 1000$\sim$2500 pixels. Here all images are resized to 100$\times$100 pixels (see Figure \ref{fig:ct}).

\subsection{Experimental settings and evaluation metrics}

Keras $^{[}$\cite{chollet2015keras}$^{]}$ is employed on an NVIDIA TITAN GPU to conduct all experiments. Adam $^{[}$\cite{Kingma2014Adam}$^{]}$ is adopted as the optimizer, with a batch of 128. To make the training procedure more stable, we train the constraining layer after training the autoencoder each $T$ epochs. The learning rate $\lambda$ is set to $10^{-3}$, the coefficient $\gamma$ is set to 10, and the update interval $T$ is set to 60.  The details of the encoder and decoder are shown in Figure~\ref{fig:structure}, with the kernel size set to 3$\times$3,  and the bottleneck feature number $n$ set to 256.  Note that the OCC task only has the single class samples for training, so the parameters are unable to be selected using the validation set. Thus, similar to previous work~$^{[}$\cite{ruff18deepsvdd,zong2018deep}$^{]}$, we train the model for $10^3$ epochs for each experiment, and report the result of the last epoch. 

The Area Under the Receiver Operating Characteristic Curve (AUC) and Area Under Precision-Recall curve (AUPR) are employed to evaluate the performance. Two types of AUPR can be calculated: taking the majority class as positive (AUPR-maj), and taking the minority class as positive (AUPR-min).

\section{Results}
\label{sec:Experiments}

In this section, we compare our algorithm to other existing one-class classification methods to show our algorithm's superior performance.

\subsection{Baseline methods}
\label{sec-comp-one}

The proposed method is compared with seven previous algorithms: OCSVM $^{[}$\cite{scholkopf2000support}$^{]}$, CAE, C-OCSVM,  SAE $^{[}$\cite{glorot2011deep}$^{]}$,  DAGMM $^{[}$\cite{zong2018deep}$^{]}$, AnoGAN $^{[}$\cite{deecke2018anomaly}$^{]}$ and D-SVDD $^{[}$\cite{ruff18deepsvdd}$^{]}$. We run each algorithm three times, and the average of the performance measures is reported. 

$\textbf{OCSVM}$ This method attempts to learn a mapping to project original samples into a hyperplane by kernel functions (e.g., linear, RBF). In our experiment, the RBF kernel is used for the OCSVM and each original image is reshaped into a vector as the input feature. 

$\textbf{CAE}$ To show the effectiveness of adding the constraint to the bottleneck feature, the performance of CAE is also reported as the baseline. The structure of CAE is the same as the autoencoder employed in ConOCC, with the reconstruction loss employed as the score of each sample. The learning rate $\lambda$ is tested in $\{10^{-3},10^{-4}\}$ for CAE, and the best result is obtained when $\lambda=10^{-4}$.

$\textbf{C-OCSVM}$ This method is a two-stage combination of CAE and OCSVM.   
First, an autoencoder (with the same structure as ConOCC) is trained to reduce the dimension, then the bottleneck feature is used to train the OCSVM. In the second stage, we use the same procedure as aforementioned in OCSVM. 

$\textbf{SAE}$ Sparse autoencoder (SAE) focuses on adding constraints on the bottleneck feature to learn sparse representations, by which features can be compressed into fewer nodes. The SAE in  $^{[}$\cite{glorot2011deep}$^{]}$ is employed with the same structure as ConOCC for this experiment. The learning rate $\lambda$ is tested in $\{10^{-3},10^{-4}\}$ for SAE, and the best performance is reported.
 
$\textbf{DAGMM}$ DAGMM employs an autoencoder (with the same structure as ConOCC) to compress data and obtain a reconstruction loss, which is further fed into a Gaussian Mixture Model (GMM). By jointly training with GMM, the autoencoder can escape from local optima and further reduce reconstruction errors of training samples.

$\textbf{AnoGAN}$ AnoGAN $^{[}$\cite{schlegl2017unsupervised}$^{]}$ is a GAN-based method for medical image classification. After training a GAN with samples from the majority class, the model can generate a latent code for a given testing image, then the difference between the generated and original image is calculated as the loss. Considering the sample number of our datasets, here the model provided by the original paper is used along with data augmentation.

$\textbf{D-SVDD}$ D-SVDD focuses on mapping samples into a hypersphere using a feature extraction network. Here the encoder of ConOCC is used as the basic model, and a SVDD-based loss layer is integrated with default parameters for this experiment.

\begin{table}[ht]
\centering
\caption{Comparison of ConOCC to seven other previous algorithms.}
\label{tab:res_all}
\setlength{\tabcolsep}{1mm}{
\begin{tabular}{c c| c c c c c c c c c c}
\toprule[1pt]
Data&Method&\small{OCSVM}&\small{COCSVM}&\small{DAGMM}&\small{D-SVDD}&\small{AnoGAN}&\small{SAE}&\small{CAE}&\small{OURS}\\
\midrule 
\multirow{3}{*}{\small{MRI}}&\multirow{1}{*}{\shortstack{AUC}} 
&0.802&0.844&0.629&0.652&0.613&0.854&0.782&\textbf{0.872}\\
\cline{2-10}
&\multirow{1}{*}{\shortstack{\small{AUPR-maj}}} 
&0.831&0.866&0.545&0.675&0.611&0.855& 0.825&\textbf{0.879}\\
\cline{2-10}
&\multirow{1}{*}{\shortstack{\small{AUPR-min}} }
&0.784&0.826&0.692&0.614&0.590&0.859&0.713&\textbf{0.871}\\
\toprule[0.9pt] 
\multirow{3}{*}{\shortstack{\small{FFDM}}}&\multirow{1}{*}{\shortstack{AUC}} 
&0.771&0.710&0.563&0.622&0.676&0.779&0.675&\textbf{0.791}\\
\cline{2-10}
&\multirow{1}{*}{\shortstack{\small{AUPR-maj}} }
&0.830&0.744&0.563&0.640&0.686&0.837&0.720&\textbf{0.839}\\
\cline{2-10}
&\multirow{1}{*}{\shortstack{\small{AUPR-min}} }
&0.700&0.677&0.660&0.590&0.681&0.701&0.622&\textbf{0.710}\\
\toprule[0.9pt] 
\multirow{3}{*}{\shortstack{\small{ChestX}}}&\multirow{1}{*}{\shortstack{AUC}} 
&0.548&0.587&0.564&0.579&0.594&0.603&0.576&\textbf{0.624}\\
\cline{2-10}
&\multirow{1}{*}{\shortstack{\small{AUPR-maj}}} 
&0.449&0.413&0.423&0.422&0.446&0.435&0.394&\textbf{0.519}\\
\cline{2-10}
&\multirow{1}{*}{\shortstack{\small{AUPR-min}}} 
&\textbf{0.756}&0.674&0.694&0.684&0.702&0.748&0.675&0.739\\

\toprule[0.9pt] 
\end{tabular}}
\end{table}
 
 \begin{table}[ht]
\centering
\caption{Comparison of ConOCC to previous algorithms when using cross-validation.}

\label{tab:res_cv}
\setlength{\tabcolsep}{1mm}{
\begin{tabular}{c c| c c c c c c c c c c}
\toprule[1pt]
Data&Method&\small{OCSVM}&\small{COCSVM}&\small{DAGMM}&\small{D-SVDD}&\small{AnoGAN}&\small{SAE}&\small{CAE}&\small{OURS}\\
\midrule 
\multirow{6}{*}{\small{MRI}}&\multirow{1}{*}{\shortstack{AUC}} 
&0.807&0.811&0.707&0.687&0.634&0.843&0.760&\textbf{0.868}\\
&\multirow{1}{*}{\shortstack{$\pm$std}} 
&0.025&0.133&0.062&0.091&0.030&0.026&0.006&0.022\\
\cline{2-10}
&\multirow{1}{*}{\shortstack{\small{AUPR-maj}}}
&0.834&0.798&0.660&0.503&0.578&0.813&0.785&\textbf{0.865}\\
&\multirow{1}{*}{\shortstack{$\pm$std}} 
&0.016&0.166&0.021&0.208&0.079&0.041&0.011&0.020\\
\cline{2-10}
&\multirow{1}{*}{\shortstack{\small{AUPR-min}}}
&0.793&0.817&0.777&0.465&0.674&0.825&0.726&\textbf{0.878}\\
&\multirow{1}{*}{\shortstack{$\pm$std}}
&0.036&0.110&0.036&0.139&0.023&0.016&0.011&0.024\\
\toprule[0.9pt] 

\multirow{6}{*}{\small{FFDM}}&\multirow{1}{*}{\shortstack{AUC}} 
&0.772&0.683&0.607&0.662&0.705&0.786&0.650&\textbf{0.827}\\
&\multirow{1}{*}{\shortstack{$\pm$std}} 
&0.069&0.042&0.109&0.105&0.077&0.062&0.042&0.044\\
\cline{2-10}
&\multirow{1}{*}{\shortstack{\small{AUPR-maj}} }
&0.788&0.663&0.656&0.669&0.448&0.790&0.627&\textbf{0.837}\\
&\multirow{1}{*}{\shortstack{$\pm$std}} 
&0.075&0.054&0.120&0.090&0.152&0.069&0.057&0.051\\
\cline{2-10}
&\multirow{1}{*}{\shortstack{\small{AUPR-min}}}
&0.747&0.674&0.710&0.642&0.479&0.755&0.646&\textbf{0.799}\\
&\multirow{1}{*}{\shortstack{$\pm$std}}
&0.076&0.051&0.057&0.101&0.150&0.084&0.079&0.065\\
\toprule[0.9pt] 

\multirow{6}{*}{\small{ChestX}}&\multirow{1}{*}{\shortstack{AUC}} 
&0.580&0.613&0.542&0.562&0.608&0.588&0.599&\textbf{0.616}\\
&\multirow{1}{*}{\shortstack{$\pm$std}} 
&0.037&0.058&0.052&0.065&0.049&0.050&0.049&0.015\\
\cline{2-10}
&\multirow{1}{*}{\shortstack{\small{AUPR-maj}} }
&0.522&0.582&0.501&0.461&0.410&0.551&0.552&\textbf{0.597}\\
&\multirow{1}{*}{\shortstack{$\pm$std}} 
&0.063&0.064&0.178&0.069&0.021&0.074&0.061&0.002\\
\cline{2-10}
&\multirow{1}{*}{\shortstack{\small{AUPR-min}}}
&0.613&0.605&0.617&\textbf{0.671}&0.465&0.607&0.638&0.619\\
&\multirow{1}{*}{\shortstack{$\pm$std}}
&0.020&0.063&0.092&0.057&0.035&0.0432&0.023&0.021\\
\toprule[0.9pt] 
\end{tabular}}
\end{table}

The average results of running the experiment three times are reported in Table~\ref{tab:res_all}; the average results when using cross-validation are reported in Table~\ref{tab:res_cv}. Given that the proportion of the minority class samples in the three datasets is different, we use varying numbers of folds for different datasets in cross-validation. Specifically, for MRI, FFDM, and ChestX datasets, we use 1020, 200, and 1185 majority class samples for training, and the rest majority and minority class samples for testing.  In the testing set of each fold, the minority class samples are the same while the majority class samples are changed 
alternately, thus the cross-validation fold number is 2, 5, and 4, respectively. From these two tables we have the following observations.

For models that focus on compactness, we observe that OCSVM and COCSVM are two relatively effective methods among all datasets. The reason could be that the kernel-based methods are effective in small-scale datasets$^{[}$\cite{scholkopf2000support,ruff18deepsvdd}$^{]}$.

For autoencoder-like models that focus on descriptiveness, we can see that CAE and SAE outperform other deep learning-based methods, indicating the reconstruction-based method may be more suitable for learning features of medical images. Also, SAE outperforms CAE, indicating that adding the constraint on the bottleneck features is useful for learning better representations. Autoencoder-like models neglect the compactness of the features. These models are trained on reconstructing inputs, and they tend to converge to a local optima for the purpose of reconstruction only, making their performances limited for anomaly detection.

Our method outperforms all the autoencoder-like models in the three datasets. Specifically, ConOCC gains a better performance than CAE, indicating that by adding the compact constraint, ConOCC can learn more distinguishable features of the given class and the decoder can further amplify the difference of different classes. ConOCC also outperforms SAE, indicating that the compactness constraint is better than the sparse constraint.

By jointly optimizing the two loss objectives, ConOCC can  simultaneously minimize the intra-class distance and maximize the inter-class distance. Our method outperforms the other compared methods in all the tested datasets, showing that learning both compact and descriptive features together obtains a better feature representation compared to methods that only focus on one aspect.

We notice the AUCs of all methods on the ChestX dataset are relatively low. This is probably because it is hard to learn representations on large-size images, especially when the training set is small.

\section{Discussion}
In this section, we discuss the following two aspects: 
1) analyzing the influence of coefficients to show the robustness of the proposed method (Section~\ref{sec-comp-para}); 2)  visualizing features to show that the proposed method can learn a compact hypersphere (Section~\ref{sec-visualization}).

\subsection{Robustness analysis}
\label{sec-comp-para}

\begin{figure*}[ht!]
\centering
\includegraphics[scale=0.32]{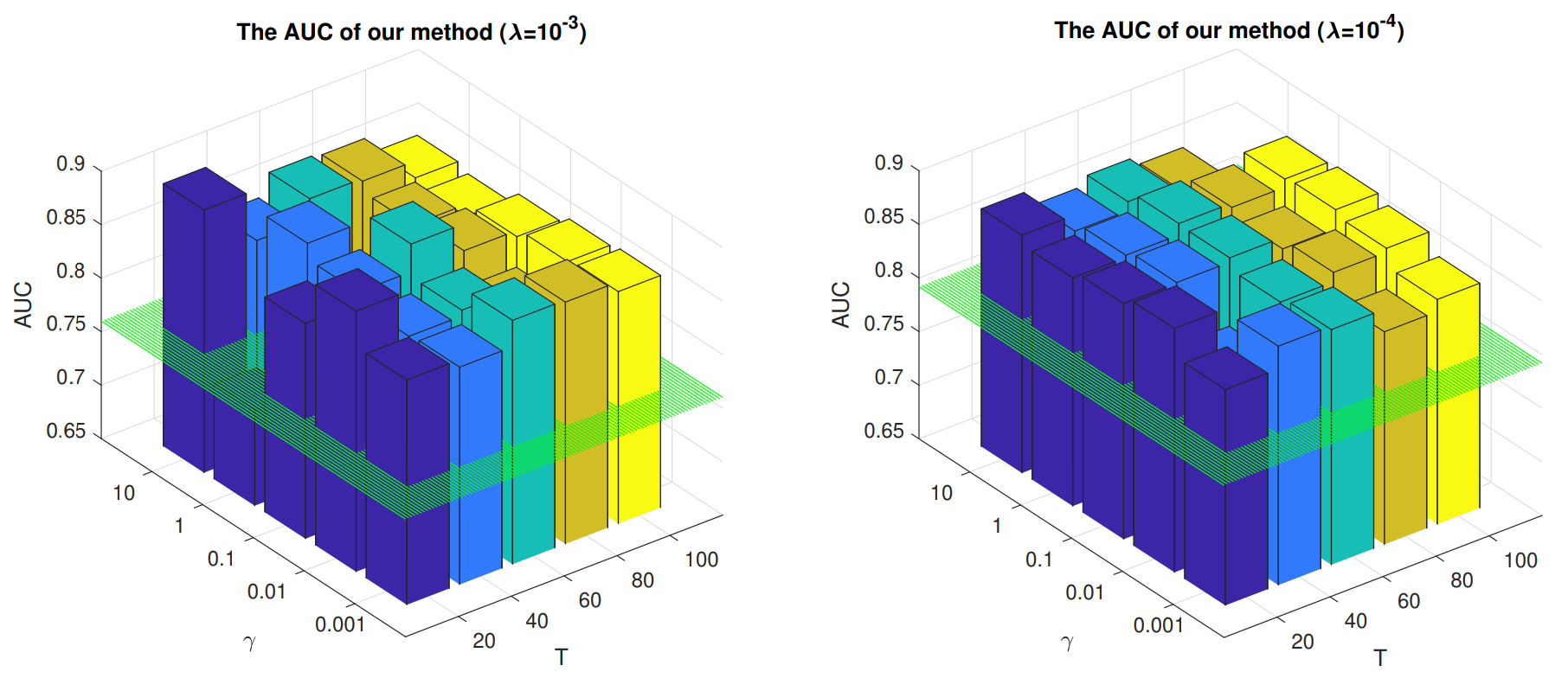}
\caption{\label{fig:robust} The influence of $\lambda$, $\gamma$ and interval T. The green plane is the result of CAE in corresponding learning rate.}
\end{figure*}  

To evaluate the influence of the $\lambda$, $\gamma$ and $T$, we take the breast MRI dataset as an example and report AUCs when setting $\lambda\in\{10^{-3}, 10^{-4}\}$, $\gamma\in$\{10$^{-2}$, 10$^{-1}$,  $\cdots$, $10^{1}$\}, and $T\in\{20, 40, \cdots, 100\}$. In Figure~\ref{fig:robust}, the green plane with the dotted line is the result of CAE. The histogram is the result of ConOCC in different settings (run one time). As shown in Figure~\ref{fig:robust}, we can see that CAE is influenced by the $\lambda$ significantly, whereas ConOCC is robust to $\lambda$. The proposed method can keep high and stable performance in a majority of $\gamma$ and $T$.

\begin{figure}[ht!]
\centering
\includegraphics[scale=0.35]{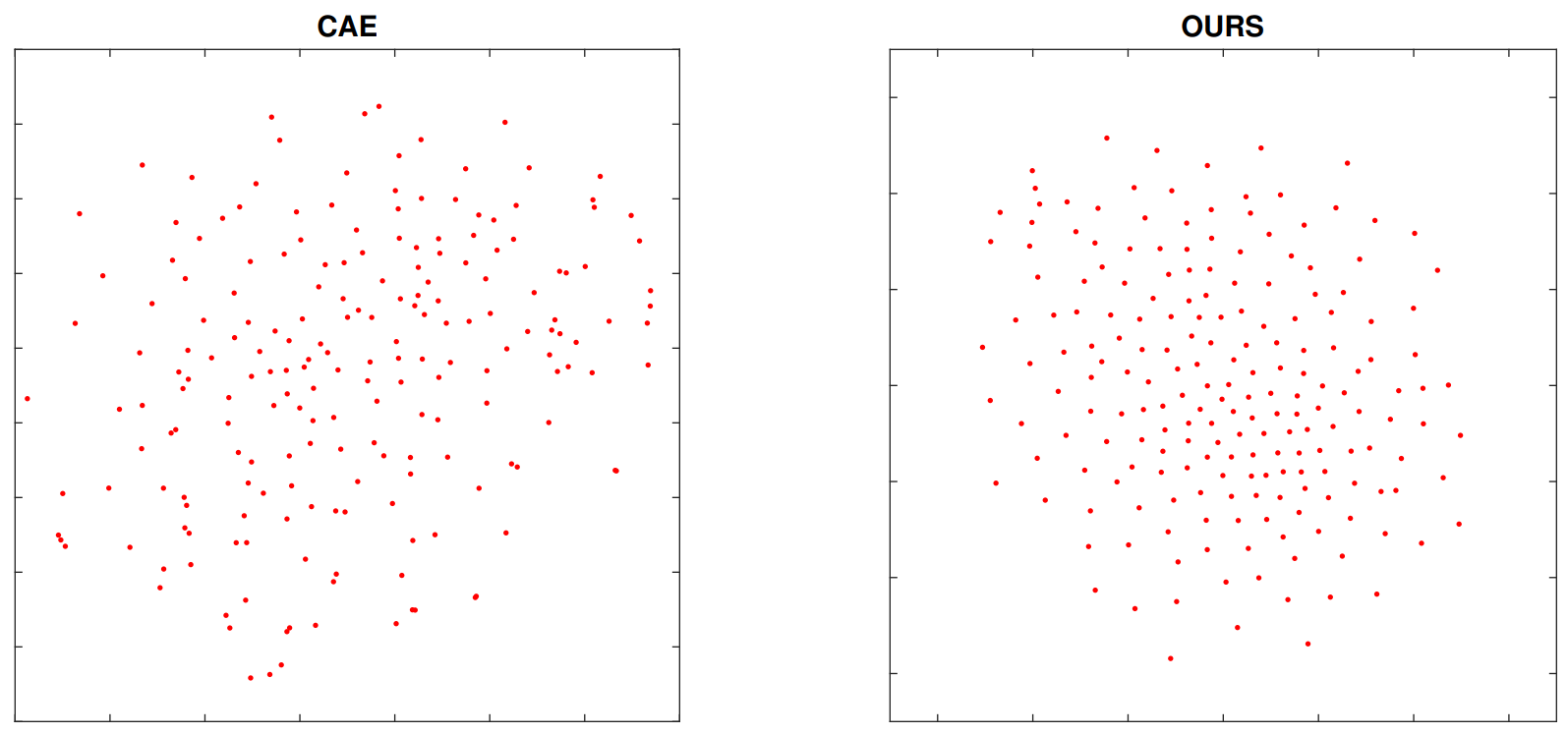}
\caption{\label{fig:tsne} Visualized features of training samples for CAE and ConOCC on the FFDM dataset.}
\end{figure}

\subsection{Visualization of features}
\label{sec-visualization}

We take the FFDM dataset as an example to illustrate the distribution of the bottleneck features. Specifically, we extract features from the training set and visualize them by reducing into two dimensions with $t$-SNE $^{[}$\cite{maaten2008visualizing,wattenberg2016use}$^{]}$. As shown in Figure~\ref{fig:tsne}, the proposed method can learn a more compact hypersphere than CAE, and the sample density increases when samples are close to the center point. This indicates that the constraining layer can generate a compact hypersphere. 

\section{Conclusions}
\label{Conclusion}

To address the data imbalance problem in medical image classification, a framework is proposed to learn constrained features by integrating a constraining layer into an autoencoder. The proposed method can jointly learn a compact while descriptive feature space for the given class samples, making samples belonging to different classes more distinguishable. The proposed method is evaluated on three clinical datasets for three detection tasks, and it exhibits improved performance compared to previous methods. In the future, we will continue to investigate this framework by integrating potentially other constraints on the bottleneck features to further improve performance.

\section*{Acknowledgement}
This project was supported in part by a Developmental Pilot Award of the Pittsburgh Center for AI Innovation in Medical Imaging and the associated Pitt Momentum Funds through a Scaling grant from the University of Pittsburgh (2020). We gratefully acknowledge the support of NVIDIA Corporation for the donation of the Titan X Pascal GPU for our research. Most of the work was conducted when the first author was a visiting student in the University of Pittsburgh, with no funding support provided to the first author.

\section*{Disclosure}
Dr. Shandong Wu is a scientific consultant and stockholder of COGNISTX, Inc. Dr. Shandong Wu has a research grant funded by Amazon. All other authors have no conflicts of interests to disclose.

\bibliography{ConOCC}

\begin{thebibliography}{10}

\bibitem{khan2014one}
S.~S. Khan and M.~G. Madden, ``One-class classification: taxonomy of study and
  review of techniques,'' {\em The Knowledge Engineering Review} {\bf 29}(3),
  345--374  (2014).

\bibitem{chalapathy2019deep}
R.~Chalapathy and S.~Chawla, ``Deep learning for anomaly detection: A survey,''
  {\em arXiv preprint arXiv:1901.03407} {\bf 1}  (2019).

\bibitem{ieracitano2019convolutional}
C.~Ieracitano, N.~Mammone, A.~Bramanti, {\em et~al.}, ``A convolutional neural
  network approach for classification of dementia stages based on 2d-spectral
  representation of eeg recordings,'' {\em Neurocomputing} {\bf 323}, 96--107
  (2019).

\bibitem{2018Learning}
P.~Perera and V.~M. Patel, ``Learning deep features for one-class
  classification,'' {\em IEEE Transactions on Image Processing} {\bf PP}(99)
  (2018).

\bibitem{Tax2004svdd}
D.~M. Tax and R.~P. Duin, ``Support vector data description,'' {\em Machine
  Learning} {\bf 54}(1), 45--66  (2004).

\bibitem{scholkopf2000support}
B.~Sch{\"o}lkopf, R.~C. Williamson, A.~J. Smola, {\em et~al.}, ``Support vector
  method for novelty detection,'' in {\em Advances in Neural Information
  Processing Systems},  582--588  (2000).

\bibitem{chen2018diverse}
B.~Chen, L.~Wang, J.~Sun, {\em et~al.}, ``Diverse lesion detection from retinal
  images by subspace learning over normal samples,'' {\em Neurocomputing} {\bf
  297}, 59--70  (2018).

\bibitem{dreiseitl2010outlier}
S.~Dreiseitl, M.~Osl, C.~Scheibb{\"o}ck, {\em et~al.}, ``Outlier detection with
  one-class svms: an application to melanoma prognosis,'' in {\em AMIA Annual
  Symposium Proceedings},   {\bf 10}, 172  (2010).

\bibitem{guo2011tumor}
L.~Guo, L.~Zhao, Y.~Wu, {\em et~al.}, ``Tumor detection in mr images using
  one-class immune feature weighted svms,'' {\em IEEE Transactions on
  Magnetics} {\bf 47}(10), 3849--3852  (2011).

\bibitem{ruff18deepsvdd}
L.~Ruff, R.~A. Vandermeulen, N.~G{\"o}rnitz, {\em et~al.}, ``Deep one-class
  classification,'' in {\em Proceedings of the 35th International Conference on
  Machine Learning},   {\bf 80}, 4393--4402  (2018).

\bibitem{yosinski2014transferable}
J.~Yosinski, J.~Clune, Y.~Bengio, {\em et~al.}, ``How transferable are features
  in deep neural networks?,'' in {\em Advances in Neural Information Processing
  Systems},  3320--3328  (2014).

\bibitem{zong2018deep}
B.~Zong, Q.~Song, M.~R. Min, {\em et~al.}, ``Deep autoencoding gaussian mixture
  model for unsupervised anomaly detection,'' in {\em International Conference
  on Learning Representations},  OpenReview.net  (2018).

\bibitem{deecke2018anomaly}
L.~Deecke, R.~A. Vandermeulen, L.~Ruff, {\em et~al.}, ``Anomaly detection with
  generative adversarial networks,'' {\em Machine Learning and Knowledge
  Discovery in Databases} {\bf 11051}, 3--17  (2018).

\bibitem{zhou2017anomaly}
C.~Zhou and R.~C. Paffenroth, ``Anomaly detection with robust deep
  autoencoders,'' in {\em International Conference on Knowledge Discovery and
  Data Mining},  665--674, ACM  (2017).

\bibitem{Goodfellow2014Generative}
I.~J. Goodfellow, J.~Pouget-Abadie, M.~Mirza, {\em et~al.}, ``Generative
  adversarial networks,'' {\em Advances in Neural Information Processing
  Systems} {\bf 3}, 2672--2680  (2014).

\bibitem{tang2019chest}
Y.~Tang, ``Deep adversarial one-class learning for normal and abnormal chest
  radiograph classification,'' in {\em SPIE},   {\bf 10950}  (2019).

\bibitem{schlegl2017unsupervised}
T.~Schlegl, P.~Seeb{\"o}ck, S.~M. Waldstein, {\em et~al.}, ``Unsupervised
  anomaly detection with generative adversarial networks to guide marker
  discovery,'' in {\em International Conference on Information Processing in
  Medical Imaging},  146--157, Springer  (2017).

\bibitem{fan2020robust}
J.~Fan, Q.~Zhang, J.~Zhu, {\em et~al.}, ``Robust deep auto-encoding gaussian
  process regression for unsupervised anomaly detection,'' {\em Neurocomputing}
  {\bf 376}, 180--190  (2020).

\bibitem{wei2018anomaly}
Q.~Wei, Y.~Ren, R.~Hou, {\em et~al.}, ``Anomaly detection for medical images
  based on a one-class classification,'' in {\em Medical Imaging 2018:
  Computer-Aided Diagnosis},   {\bf 10575}, 105751M, International Society for
  Optics and Photonics  (2018).

\bibitem{tschannen2018recent}
M.~Tschannen, O.~Bachem, and M.~Lucic, ``Recent advances in autoencoder-based
  representation learning,'' {\em arXiv preprint arXiv:1812.05069} {\bf 1}
  (2018).

\bibitem{glorot2011deep}
X.~Glorot, A.~Bordes, and Y.~Bengio, ``Deep sparse rectifier neural networks,''
  in {\em International Conference on Artificial Intelligence and Statistics},
  315--323  (2011).

\bibitem{chollet2015keras}
F.~Chollet, ``Keras.'' \url{https://keras.io}  (2015).

\bibitem{Kingma2014Adam}
D.~P. Kingma and J.~Ba, ``Adam: {A} method for stochastic optimization,'' {\em
  CoRR} {\bf abs/1412.6980}  (2014).

\bibitem{maaten2008visualizing}
L.~v.~d. Maaten and G.~Hinton, ``Visualizing data using t-sne,'' {\em Journal
  of Machine Learning Research} {\bf 9}, 2579--2605  (2008).

\bibitem{wattenberg2016use}
M.~Wattenberg, F.~Vi{\'e}gas, and I.~Johnson, ``How to use t-sne effectively,''
  {\em Distill} {\bf 1}(10), e2  (2016).

\end{thebibliography}
\bibliographystyle{spiejour}   

\end{spacing}
\end{document}